\documentstyle[preprint,aps]{revtex}

\tighten

\newcommand{\preprintnumber}{CERN-TH/95-110}

\begin{document}
\preprint{hep-th/9509033{\hspace{8cm}}\preprintnumber/Revised}

\draft

%****************************      TITLE    *****************************

\title{Charged black points in General Relativity\\
       coupled to the logarithmic $U(1)$ gauge theory}

%****************************     AUTHORS   ****************************

\author{Harald H. Soleng\thanks{Electronic address: soleng@surya11.cern.ch}}

\address{
Theory Division, CERN, CH-1211
Geneva 23, Switzerland}

\date{July 7, 1995}

\bibliographystyle{unsrt}

\maketitle

%***************************     ABSTRACT    *****************************

\begin{abstract}%
The exact solution for a static spherically symmetric field
outside a charged point particle is found in a non-linear
$U(1)$ gauge theory with a logarithmic Lagrangian. The
electromagnetic self-mass is finite, and for a particular
relation between mass, charge, and the value of the non-linearity
coupling constant, $\lambda$, the electromagnetic contribution
to the Schwarzschild mass is equal to the total mass.
If we also require that the singularity at the origin be hidden
behind a horizon, the mass is fixed to be slightly less than
the charge. This object is a {\em black point.}
\end{abstract}


%
%***************************    TEXT       *******************************

The singularity problem of Einstein's general theory of relativity
has sometimes been regarded as a ``crisis in physics'' \cite{MTW}.
It is hard to
accept a theory in which space--time itself breaks down and where the
Riemann tensor is predicted to diverge on a
singularity which can be reached along a time-like
curve.
In general, physicists seem to have had
less trouble with analogous singularities
in gauge theories. One reason might be
that in gauge theories the diverging curvature tensors
are curvatures not of space--time but of {\em internal\/} spaces.
Yet, in a sense, these internal dimensions are just as real as
the external dimensions of everyday space--time. Therefore we
should feel just as embarrassed by these singularities as by the
singularities of Einstein's theory. In addition, near a point charge
not only does the Faraday tensor diverge, but also the electromagnetic
energy--momentum tensor blows up. Thus,
through the gravitational field equations,
such electromagnetic singularities are
also producing singularities in space--time.

Within the framework of electromagnetism, an
action for a bounded field strength was proposed
long ago
by Born and Infeld
\cite{BI:PRSLA34}.
Altshuler \cite{Altshuler:CQG90}
considered non-linear electrodynamics as a possible
mechanism for inflation, and devised a Lagrange multiplier
scheme for constructing
non-singular field theories.
This method was later
invoked to realize the
{\em limiting curvature hypothesis\/}  in
cosmological theories
\cite{MB:PRL92,BMS:PRD93}.  In two-dimensional
space--times it
has been applied both to
black holes \cite{TMB:PLB93}
and cosmological models \cite{MT:PRD95}.
But non-linear electrodynamics is not only inspired by
the desire to find non-singular
field theories; Euler and Heisenberg
\cite{HE:ZP36}
discovered that vacuum polarization
effects can be simulated classically
by a non-linear theory. Also in string
theory one has found effective actions describing
non-linear electromagnetism \cite{FT:PLB85}.

In this Brief Report I investigate
the logarithmic $U(1)$ gauge theory
which is
contained in the class of theories constructed
by Altshuler
\cite{Altshuler:CQG90}.
This particular case was omitted in the
analysis of non-linear charged black holes
carried out by de~Oliveira \cite{deOliveira:CQG94}.
While this particular theory
appears to have no direct relation to superstring
theory, it serves as a toy-model illustrating that
certain non-linear field theories can produce
particle-like solutions which can realize the
{\em limiting curvature hypothesis\/} also for gauge fields.

I shall find
the classical (non-linear)\
electromagnetic and gravitational field
for a static charged
point particle.
For the electromagnetic field there are in general two invariants which need
to be bounded: $I_1\equiv F_{\alpha\beta}F^{\alpha\beta}$ and
$I_{2}\equiv {}^{*}F_{\alpha\beta}F^{\alpha\beta}$.
For a static,
charged point particle, the latter invariant vanishes identically.
Therefore
I shall only consider $I_1$.

The action $S=\int{\cal{L}}\sqrt{-g}d^4x$ is
specified by
the Lagrangian density
\begin{equation}
{\cal{L}}=\frac{1}{16\pi\lambda}
\left[\lambda R -
\ln{(1+\lambda F_{\alpha\beta} F^{\alpha\beta})}
\right]
\label{Eq:Lagrangian}
\end{equation}
where geometrized units \cite{MTW}
with $G=c=1$ have been employed.
In these units
the constant $\lambda$ has dimension $({\mbox{length}})^2$.
The lowest-order terms of the Lagrangian are
\begin{equation}
{\cal{L}}= \frac{1}{16\pi}
\left[ R- F_{\alpha\beta} F^{\alpha\beta}
+\frac{\lambda}{2} (F_{\alpha\beta} F^{\alpha\beta})^2
+{\cal{O}}(\lambda)^2 \right].
\label{Eq:AppLagrangian}
\end{equation}
To second order, and when $I_2=0$, both
the Born--Infeld
\cite{BI:PRSLA34}
and the Euler--Heisenberg
\cite{HE:ZP36}
actions
can be represented by the logarithmic Lagrangian.
With the action (\ref{Eq:Lagrangian}), the energy--momentum tensor is
\begin{eqnarray}
8\pi T_{\mu\nu}&=&
2
 \left(1+\lambda F_{\alpha\beta} F^{\alpha\beta} \right)^{-1}
F_{\mu\rho} {F_{\nu}}^{\rho}\nonumber\\
& &{} -\frac{1}{2} g_{\mu\nu}
\frac{1}{\lambda}\ln
(1+\lambda F_{\alpha\beta} F^{\alpha\beta}).
\label{Eq:Tmunu}
\end{eqnarray}
The inhomogeneous
electromagnetic field equations
are
\begin{equation}
 \left[\left(1+\lambda F_{\alpha\beta} F^{\alpha\beta} \right)^{-1}
F^{\mu\nu}\right]_{;\nu}= 4\pi J^{\mu}.
\label{Eq:EMFEq}
\end{equation}
The homogeneous
(cyclic)
equations are identities which remain unchanged.

Let us now consider a charged point particle at rest.
Thus the space--time
metric
is given by
the spherically symmetric static metric. In Schwarzschild coordinates
the line element is \cite{rad-boost}
\begin{equation}
ds^2= C(r)^{-2} dr^2+r^2 d\theta^2+r^2 \sin^2\!\theta
\,
d\phi^2- C(r)^2 dt^2.
\label{Eq:Metric}
\end{equation}
The electromagnetic vector potential
is given by
\begin{equation}
A^{\mu}=\frac{V(r)}{C(r)}\,{\delta^{\mu}}_{4}
\end{equation}
relative to the natural orthonormal
frame
${\mbox{\boldmath{$\omega$}}}^{1}=
C(r)^{-1}{\mbox{\boldmath{$d$}}} r$,
${\mbox{\boldmath{$\omega$}}}^{2}=
r{\mbox{\boldmath{$d$}}} \theta$,
${\mbox{\boldmath{$\omega$}}}^{3}=
r \sin\theta {\mbox{\boldmath{$d$}}} \phi$,
and
${\mbox{\boldmath{$\omega$}}}^{4}=
C(r) {\mbox{\boldmath{$d$}}} t$.
The only non-vanishing
component of the Faraday tensor is
the radial component of the electric field
\begin{equation}
F_{14}=-V'(r).
\label{Eq:Faraday}
\end{equation}
With these assumptions, the electromagnetic field
equation (\ref{Eq:EMFEq}) takes the form
\begin{eqnarray}
& & 2\,V'(r) - 4\,{\lambda}\,{{V'(r)}^3} + r\,V''(r) \nonumber \\
& & \ \ +
2\,{\lambda}\,r\,{{V'(r)}^2}\,V''(r)
=0.
\end{eqnarray}
The first integral is
\begin{equation}
V'(r)=\frac{r^2-\sqrt{8\lambda Q^2+r^4}}{4\lambda Q}.
\label{Eq:Vr}
\end{equation}
By comparison with the case $\lambda=0$,
we have identified the integration constant
with the charge $Q$.
Integrating once more, we find
\begin{eqnarray}
   V(r) & = &
\frac{r^3 - r \sqrt{8\lambda Q^2 + r^4}}{
          12\lambda Q} \nonumber\\
& & {}-
   \frac{\sqrt{2} Q r
       \, {}_{\,2\!}F_{1}(\frac{1}{4},\frac{1}{2};\frac{5}{4};
        \frac{-r^4}{8\lambda Q^2})}{3\sqrt{\lambda}|Q|} + C_{1} ,
\end{eqnarray}
where ${}_{\,2\!}F_{1}$ is the generalized hypergeometric function
and $C_{1}$ is
a constant.

For small $r$ we find
a linear potential
\begin{equation}
V(r)= - \frac{Q}{\sqrt{2 \lambda} |Q|} r + {\cal{O}}(r)^3 .
\end{equation}
A linear ultraviolet
potential
has also
been found
\cite{SG:AoP95}
in a Kaluza--Klein model based on a five-dimensional
Lovelock theory.
Apart from an irrelevant  constant, which has been
neglected,
the Coulomb potential
is
the leading term
for large $r$:
\begin{equation}
V(r)= \frac{Q}{r} - \frac{2}{5}\frac{\lambda Q^3}{r^5} +
{\cal{O}}(1/r)^8.
\end{equation}

With this exact solution for the electromagnetic field, we can
integrate Einstein's field equations. Note that the $44$-component
of the Einstein tensor for the metric (\ref{Eq:Metric})
can be written
\begin{equation}
G_{44}=\frac{1}{r^2}\, \frac{d}{dr}\left[ r\left( 1-C(r)^2\right)\right].
\end{equation}
Consequently, Einstein's
field equations reduce to
\begin{equation}
C(r)^2=
1-\frac{8\pi}{r} \int_{0}^{r} r'^2 T_{44}(r') dr'
-\frac{2M_{0}}{r}.
\end{equation}
The last term is a contribution to the Schwarzschild mass coming from
a source at the origin. From now on we shall set $M_0=0$.
We note however that a non-zero $M_{0}$ generates a more Schwarzschild-like
space--time structure.
The explicit form of $T_{44}$ can now be computed from
Eqs.~(\ref{Eq:Tmunu}) and (\ref{Eq:Faraday}). Despite the complexity of the
resulting integrand,
the integral can be evaluated exactly. The result (with $M_{0}=0$) is
\begin{eqnarray}
C(r)^2 &=&
  1 + {{5\,{r^2}}\over {18\,\lambda }} -
   {{5\,{\sqrt{8\,\lambda \,{Q^2} + {r^4}}}}\over {18\,\lambda }} \nonumber\\
& & {} -
   \frac{4\sqrt{2} |Q|\,
       {}_{\,2\!}F_{1}(\frac{1}{4},\frac{1}{2};\frac{5}{4};
        -\frac{r^4}{8\lambda Q^2})}
{9\sqrt{\lambda}}
\nonumber\\
& &{} -
   \frac{r^2}{6\,\lambda} \,\ln \left(\frac{-r^4 +
              \sqrt{8\lambda Q^2 r^4 + r^8}  }{
          4\lambda Q^2}\right).
\label{Eq:Aexact}
\end{eqnarray}
For small $r$ the metric coefficient is
\begin{eqnarray}
C(r)^2&= & 1 -
    \frac{\sqrt{2}\,|Q|}{\sqrt{\lambda}} +
    \frac{5 r^2}{18\,\lambda }  \nonumber\\
& & {} -
      \frac{r^2}{6\lambda} \ln \left(\frac{r^2}{
              \sqrt{2\lambda} |Q|}\right)
         + {\cal{O}}(r)^3.
\label{Eq:Asmall}
\end{eqnarray}
Even though this metric seems to be well-behaved
at the origin, there is still a curvature
singularity there; at small radii
the leading order
of the Kretschmann
invariant is
\begin{equation}
R_{\alpha\beta\gamma\delta}
R^{\alpha\beta\gamma\delta}
=\frac{8 Q^2}{\lambda r^4}
+ {\cal{O}}(1/r)^2.
\end{equation}
This singularity should, however, not come as a surprise;
we have not attempted to limit the space--time curvature.
On the other hand, this singularity is much weaker than the
singularities of the conventional Reissner--Nordstr{\"o}m
and Schwarzschild space--times.

{}From Eq.~(\ref{Eq:Asmall})
one finds that
$C(r)$ changes sign near the origin if $\lambda < 2 Q^2$.
This means that
there is a horizon at a small radius
and that the model is a black hole
if $\lambda$ is small.
If $\lambda=2 Q^2$, then the horizon is at the origin.
Such an object is a point-like black hole and we shall call these
objects {\em black points.}
This is not the first occurrence of a point-like black hole;
charged dilatonic
black holes \cite{GM:NPB88,GHS:PRD91}
with dilaton coupling constant $a>0$ also reduce to
black holes with a vanishing horizon radius in the extremal case. For
$a>1$ the dilatonic black points
behave
physically
as elementary particles
\cite{HW:NPB92}.
In the limit $\lambda\rightarrow 0$, we expect to recover
properties of the Reissner--Nordstr{\"o}m solution and
the appearance of a horizon agrees with this expectation.
There is however no inner horizon for any non-zero
$\lambda$.

At large radii we get the following asymptotic form
\begin{equation}
C(r)^2= 1-
 \frac{2^{5/4}\,\Gamma(\frac{1}{4})^2\,
      |Q|^{3/2}}{9\,\sqrt{\pi}\, \lambda^{1/4} \, r}
+\frac{Q^2}{r^2} + {\cal{O}}(1/r)^3
\end{equation}
where $\Gamma(x)$ is the {\em gamma function.}
We note that a Schwarzschild mass has been generated by the field (another
contribution to the
Schwarzschild mass term can be added by assuming a point mass
$M_{0}\neq 0$
at the origin).
It is possible that effects of this type can explain
a charged particle's
mass in terms of electromagnetic field energy.
It is therefore of interest
to see what the size of the $\lambda$ coupling
must be in order that this is the case. For a particle with charge $Q$
and rest mass $M=m_{0}$,
the electromagnetic contribution to the
Schwarzschild mass
is equal to the rest mass if
and only if
$\lambda=\lambda_{0}$,
where
\begin{equation}
\lambda_{0} =
\frac{2\, \mu_{0}^4 \,Q^6}{m_{0}^4}
\label{Eq:lambda0}
\end{equation}
and
\begin{equation}
\mu_{0}\equiv
\frac
{\Gamma(\frac{1}{4})^2}
{9\,\sqrt{\pi}}
\approx 0.824033.
\end{equation}
For $\lambda>\lambda_{0}$ there is a positive point mass
at the centre of symmetry, and if $\lambda<\lambda_{0}$, the
central mass must be negative.
A negative central mass is also found in the
pure Reissner--Nordstr{\"o}m case;
here the well-known Reissner--Nordstr{\"o}m
repulsion must be caused by a genuinely
negative gravitational mass.

The presence of a non-zero $\lambda$ implies that
the Coulomb interaction changes character at a radius $r_{\text{cr}}$
where
\begin{equation}
\left. V'(r)^2\right|_{r=r_{\text{cr}}}= 1/\lambda.
\end{equation}
Using the solution (\ref{Eq:Vr}),
we find that the critical radius is given by
\begin{equation}
r_{\text{cr}}
=\lambda^{1/4} |Q|^{1/2}.
\end{equation}
At smaller scales the electromagnetic field becomes effectively
$r$-independent.

It has long been conjectured that all or nearly all
of the mass of the lightest charged particle is of electromagnetic origin.
If we insert
the value $\lambda=\lambda_{0}$ with
$Q=e$ and
$m_{0}=m_{\text{e}}$ (the electron mass),
we find $r_{\text{cr}}\approx
3\times 10^{-13}$~cm.
This is of the same size as the classical electron radius
or around 100 MeV in energy units.
This might
look appealing, but
the model fails because
in this case
the critical value of $\lambda$ is about an order of magnitude larger than
the size of
the corresponding coupling in the Euler--Heisenberg \cite{HE:ZP36}
action.
High-precision experiments in QED rule out such a large
value of $\lambda$.

It is perhaps more natural to look for these effects
at the Planck scale \cite{Brandenberger:95}. Indeed, there is
a cosmic censorship argument that leads to
$\lambda$ at such a large scale.
In addition to the requirement that the whole mass is
generated by the field, one can also demand that the
space--time singularity at
the origin should not be naked.
{}From the small-distance behaviour of the
metric (\ref{Eq:Asmall}), one finds that $r=0$ is a horizon if
\begin{equation}
|Q|=\sqrt{\lambda/2}.
\label{Eq:Qfound}
\end{equation}
This extremal (in the sense that it is on
the verge of becoming a naked singularity) solution describes a black point.
Since the
``point gravity'' (the analogue of the ``surface gravity'' of a black hole)
and the horizon area vanish, both
the Hawking temperature and the entropy formally vanish,
but for these objects the statistical description is
probably inappropriate \cite{Petal:MPLA91}.
If we combine the constraint (\ref{Eq:Qfound}) with
Eq.~(\ref{Eq:lambda0}), we get a unique value for the mass:
\begin{equation}
m_{0}=\mu_{0} Q.
\end{equation}
Using $Q=e/3$ by analogy with quarks, as suggested by
Rosen \cite{Rosen:FP94}, gives
$m_{0}\approx 5.1\times 10^{-7}\; {\mbox{g}} =
2.9 \times 10^{17}\; {\mbox{GeV}}$.

The Planck scale plays a r{\^{o}}le in low-energy physics; in
geometrized units the
elementary unit of charge is $e=\sqrt{\alpha}\ell_{\text{P}}$.
Since charge implies an electromagnetic field, and since this field
must have an energy that is equivalent to a rest mass, one should
naturally expect any charged particle to have a mass not much smaller
than the
Planck mass. Nature is different.
The
great puzzle it presents to us is
not why the electron has a mass but why its mass is so small.
The solution to this problem must be sought at the Planck
scale.

\acknowledgements
I wish to thank
R. Brandenberger, S. Deser, and F. Wilczek for
comments and for
pointing out useful references.
The computations in this Brief Report were carried out with help of
the
{\em Mathematica\/} package {\sc Cartan} \cite{Cartan}.

%+++++++++++++++++++++ other ENDNOTES and CITATIONS +++++++++++++


\begin{thebibliography}{100}
\bibitem{rad-boost} Since the energy--momentum tensor is
radially boost-invariant,
we need only one independent function in the metric.
\end{thebibliography}
\end{document}